\documentclass[runningheads]{llncs}

\def\BibTeX{{\rm B\kern-.05em{\sc i\kern-.025em b}\kern-.08em
    T\kern-.1667em\lower.7ex\hbox{E}\kern-.125emX}}
\usepackage{graphicx}
\usepackage{mathtools}
\usepackage{lipsum} 
\usepackage[numbers]{natbib}

\usepackage{booktabs}

\usepackage[ruled]{algorithm2e}

\usepackage{booktabs}
\usepackage{url}
\usepackage{color}
\usepackage{multirow}
\usepackage{enumitem}
\usepackage{array}
\usepackage{siunitx}
\usepackage[labelfont={bf}]{caption}
\usepackage{balance}
\usepackage{array} 
\usepackage{graphicx}
\usepackage{todonotes}
\usepackage{subfloat}
\usepackage{subfig}
\usepackage[nolist,nohyperlinks]{acronym}
\usepackage{footnote}
\makesavenoteenv{tabular}
\usepackage[perpage]{footmisc}
\usepackage[english]{babel}
\usepackage[autostyle]{csquotes}
\acrodef{GI}{gastrointestinal}
\acrodef{CNN}{Convolutional Neural Network}
\acrodef{ROI}{Regions of Interest}
\acrodef{AI}{Artificial Intelligence} 
\acrodef{CAI}{Computer Assisted Intervention} 
\acrodef{ML}{Machine Learning}
\acrodef{IOU}[IoU]{Intersection over Union}
\acrodef{FCM}{Fuzzy C-mean clustering}
\begin{document}
\title{Kvasir-SEG: A Segmented Polyp Dataset}
\author{Debesh Jha\inst{1,2}\and 
Pia H. Smedsrud\inst{1,3,4}\and
Michael A. Riegler\inst{1,7}\and 
P\aa l Halvorsen\inst{1,6} \and
Thomas de Lange\inst{4,5}\and
Dag Johansen\inst{2}\and
H\aa vard D. Johansen\inst{2}
} 
\authorrunning{Jha et al.}
\institute{SimulaMet, Norway \and UIT The Arctic University of Norway  \and  Augere Medical AS, Norway \and University of Oslo, Norway \and
Oslo University Hospital, Norway \and Oslo Metropolitan University, Norway \and Kristiania University College, Norway\\
\email{debesh@simula.no}
}%
\maketitle              
%
\begin{abstract}
Pixel-wise image segmentation is a highly demanding task in medical-image analysis. In practice, it is difficult to find annotated medical images with corresponding segmentation masks. In this paper, we present  Kvasir-SEG: an open-access dataset of gastrointestinal polyp images and corresponding segmentation masks, manually annotated by a medical doctor and then verified by an experienced gastroenterologist. Moreover, we also generated the bounding boxes of the polyp regions with the help of segmentation masks. We demonstrate the use of our dataset with a traditional segmentation approach and a modern deep-learning based \ac{CNN} approach. The dataset will be of value for researchers to reproduce results and compare methods.
By adding segmentation masks to the Kvasir dataset, which only provide frame-wise annotations, we enable multimedia and computer vision researchers to contribute in the field of polyp segmentation and automatic analysis of colonoscopy images.  

\end{abstract}
\section{Introduction}
\vspace{-1mm}
Colorectal cancer is the second most common cancer type among women and third most common among men~\cite{torre2015global}. Polyps are precursors to colorectal cancer and therefore important to detect and remove at an early stage. 
Polyps are found in nearly half of the individuals at age 50 that undergo a colonoscopy screening, and their frequency increase with age~\cite{rundle2008colonoscopic}. 
Polyps are abnormal tissue growth from the mucous membrane, which is lining the inside of the GI tract, and can sometimes be cancerous. Colonoscopy is the gold standard for detection and assessment of these polyps with subsequent biopsy and removal of the polyps. Early disease detection has a huge impact on survival from colorectal cancer~\cite{haggar2009colorectal}. In addition, several studies show that polyps are often overlooked during colonoscopies, with polyp miss rates of 14 to 30\% depending on type and size of the polyps~\cite{van2006polyp}. Increasing the detection of polyps has been shown to decrease risk of colorectal cancer~\cite{kaminski2017increased}.
Thus, automatic detection of more polyps at an early stage can play a crucial role in prevention and survival from colorectal cancer. This is the main motivation behind the development of a polyp segmentation dataset.

Image segmentation is the technique of dividing images into meaningful \acp{ROI} that are simple to analyze and interpret. Further research in medical image segmentation can assist processes such as monitoring pathology, improving the diagnostic ability by increasing accuracy, precision, and reducing manual intervention~\cite{pham2000current}. In particular, for \acp{CAI}, pixel-wise semantic segmentation methods have a huge potential to become part of fast, accurate and cost-effective systems. 

\begin{figure}[bt]
     \subfloat{%
       \includegraphics[width=0.33\textwidth]{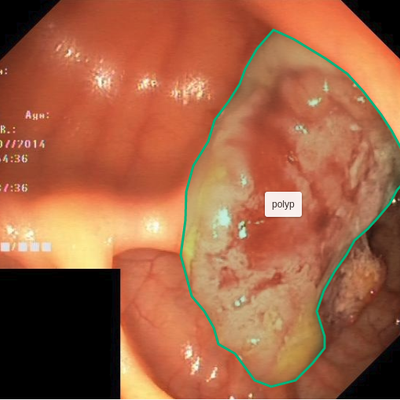}
     }
     \hfill
     \subfloat{%
       \includegraphics[width=0.33\textwidth]{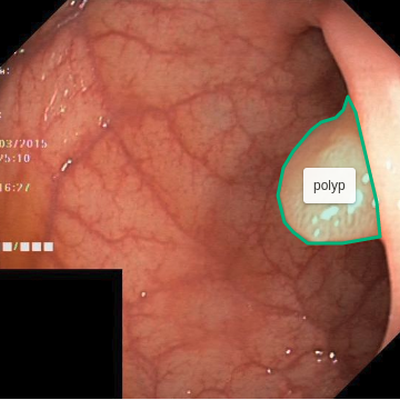}
       }
       \subfloat{%
       \includegraphics[width=0.33\textwidth]{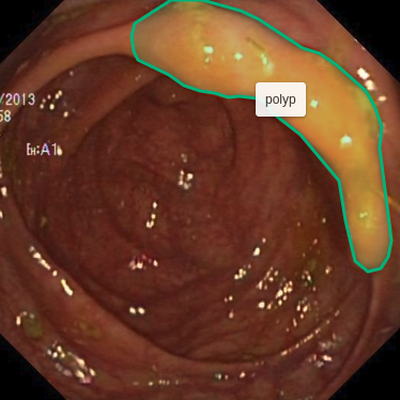}
     }
     \caption{Example frames from the Kvasir dataset where we additionally have marked the polyp tissue with green outlines.}
     \label{fig:mask_extraction}
     \vspace{-4mm}
   \end{figure}

The goal of image segmentation is to assign a label to each pixel of the image so the pixels with the same label share specific characteristics, e.g., the pixels covered by the outline in the Figure~\ref{fig:mask_extraction} show a polyp. Manual segmentation by physicians is still the gold standard for most of the medical imaging modalities, for example, Magnetic Resonance Imaging for evaluating hippocampal atrophy in Alzheimer's Disease~\cite{boccardi2011survey} and tumor segmentation of glioma~\cite{visser2019inter}. However, manual image segmentation is tedious, time-consuming, and subject to physician's bias and inter-observer variation. Therefore, there is a need for an automated and efficient image segmentation technique. 
Methods for automated and efficient image segmentation are difficult to develop as state-of-the art machine learning methods often require large number of annotated and labelled quality data, which is difficult to obtain in this field. Annotating medical data such as polyp images manually requires a lot of time and effort. It also requires medical experts, gastroenterologists in our case, which can be expensive and inaccessible. Also, there are problems related to the collection of medical images, with concern to privacy and security for patients and hospitals. \citet{riegler2016multimedia} have raised several open questions about the medical world that need to be addressed, where they emphasized the need for test datasets, including annotations and ground truth that meet current medical standards. Although there are a few available datasets, open-access datasets for comparable evaluations are missing in this field. We therefore provide the Kvasir-SEG dataset and propose a baseline model for evaluation.

The main contribution of this paper are as follows:

\begin{enumerate}
\item We extend the Kvasir dataset~\cite{kvasir} with polyp images along with their corresponding segmentation masks and bounding boxes. The \acp{ROI} are the pixels depicting polyp tissue. These are represented by a white foreground in the segmentation masks. The \acp{ROI} are generated from manual annotations verified by an experienced gastroenterologist. The bounding boxes are the set of coordinates that encloses the polyp regions. The Kvasir-SEG dataset is made publicly available and open access.

\item In this article, we include a first attempt to use the Kvasir-SEG dataset for pixel-wise semantic segmentation based analysis. For the experiment, we have used \ac{FCM}~\cite{cai2007fast} and Deep Residual U-Net (ResUNet)~\cite{zhang2018road} architecture. We achieved promising results with our proposed methods when evaluated on the same dataset. We evaluated the proposed method using Dice Coefficients and mean \ac{IOU}. These metrics were selected for fair comparison, and we encourage the use of these and similar metrics in future work on the dataset. The promising results demonstrated in this paper serves as baseline and motivation for further research and evaluation done on the same dataset.

\item Multiple datasets are prerequisites for comparing computer vision based algorithms, and this dataset is useful both as a training dataset or as a validation dataset. This dataset can assist the development of state-of-the-art solutions on images captured by colonoscopes. Further research in this field has the potential to help reduce the polyp miss rate and thus improve examination quality. 
\end{enumerate}

This paper is organized as follows: Section~\ref{sec:Related_work} discusses related datasets. We discuss the Kvasir-SEG dataset in Section~\ref{sec:Dataset}. In Section~\ref{sec:Suggested_Metrics}, we define the suggested metrics for the segmentation of polyps. Section~\ref{sec:Baseline} describes the baseline experiments, results and discussion. We conclude our work and give future directions in Section~\ref{sec:conclusion}. 
%
\section{Related Work}
\vspace{-1mm}
\label{sec:Related_work}
There are only few available polyp datasets that consist of ground truth and corresponding segmentation mask. These are  CVC-ColonDB~\cite{tajbakhsh2015automated}, ASU-Mayo Clinic Colonoscopy Video {\textcopyright} Database~\cite{bernal2012towards}, ETIS-Larib Polyp DB~\cite{silva2014toward}, and CVC-Clinic DB~\cite{bernal2015wm}.

CVC-ColonDB~\cite{tajbakhsh2015automated} is the second largest database available and consists of annotated video sequences from colonoscopy videos. From 15 short colonoscopy sequences, 1200 image frames are extracted. Out of these images, only 300 frames are annotated. These annotated frames were specifically chosen to maximize the the visual differences between them. Use of the CVC-ColonDB requires registration.

The ASU-Mayo Clinic Colonoscopy Video {\textcopyright} Database~\cite{bernal2012towards} is the first and largest available dataset captured using standard colonoscopes. The training dataset consists of 18,781 frames extracted from 20 short videos. Of these, there are 10 videos of polyps (positive) and 10 videos without polyps (negative). Ground truth and its corresponding segmentation masks are provided with the more than 3,500 frames showing polyps. For testing, 18 videos without ground truth are included. The images in the dataset are very similar to each other, which raise the problem of overfitting~\cite{kvasir}. The ASU-Mayo Clinic Colonoscopy database is copyrighted, and is only available through direct contact with the administrators at Arizona State University. 


The ETIS-Larib Polyp DB~\cite{silva2014toward} consists of 196 frames of polyps extracted from colonoscopy videos and their corresponding masks. This database is available through registration. 

The CVC-Clinic DB~\cite{bernal2015wm} consists of 612 image frames extracted from 29 different colonoscopy sequences and their corresponding ground truth as segmentation masks. The use of this database is public and open access. 

The CVC-Clinic DB, ETIS-Larib and ASU-Mayo Clinic Colonoscopy Video DB were used at Medical Image Computing and Computer Assisted Intervention (MICCAI) 2015 Automatic Polyp Detection in Colonoscopy Videos Sub-Challenge. More details about the dataset and competition can be found in the paper by \citet{bernal2017comparative}.

The literature review shows that there are few available datasets. However, an open-access dataset for comparable evaluation is missing in this field. Therefore, it was a logical next step to extend the Kvasir dataset with segmentation masks. The presented data and baseline work can be a important source for addressing the problem of standard datasets for evaluation, and help develop robust and efficient systems.
\section{The Kvasir-SEG dataset}
\vspace{-1mm}
\label{sec:Dataset}

The Kvasir-SEG dataset is based on the previous Kvasir~\cite{kvasir} dataset, which is the first multi-class dataset for \ac{GI} tract disease detection and classification.

\subsection{The original Kvasir dataset}
The original Kvasir dataset~\cite{kvasir} comprises 8,000 \ac{GI} tract images from 8 classes where each class consists of 1000 images. We replaced the $13$ images from the polyp class with new images to improve the quality of the dataset. These images were collected and verified by experienced gastroenterologists from Vestre Viken Health Trust in Norway. The  classes include anatomical landmarks, pathological findings and endoscopic procedures. A more detailed explanation about each image classes, the data collection procedure and the dataset details can be found in~\cite{kvasir}.

\begin{figure} [t]
    \centering
    \includegraphics[height=3.06cm]{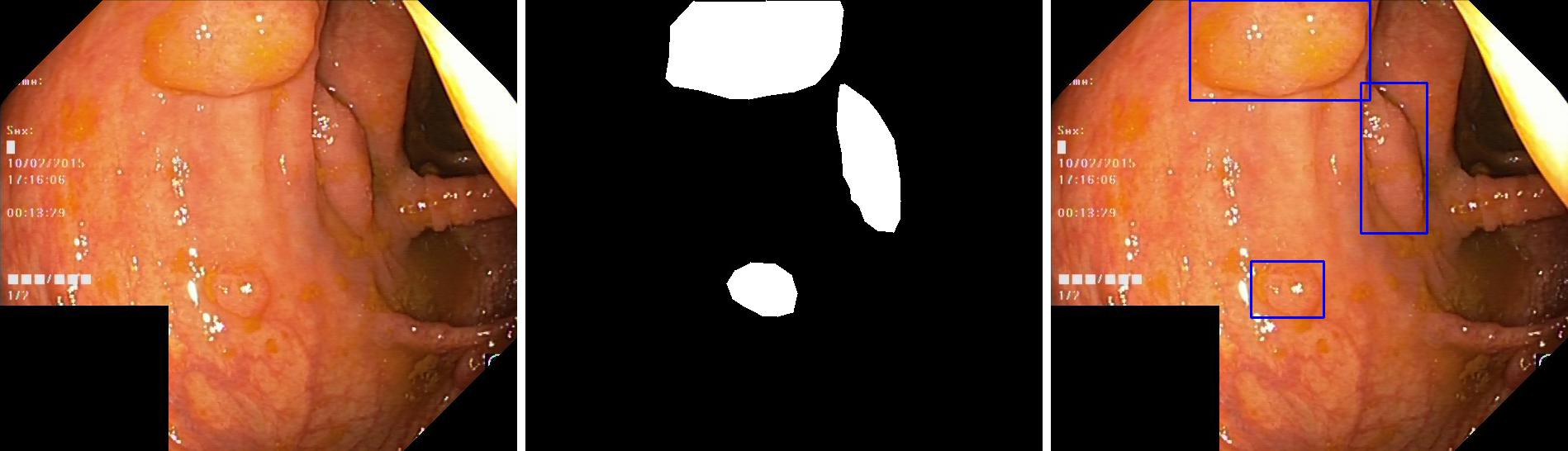}\\
    \vspace{0.5mm}
    \includegraphics[height=3cm]{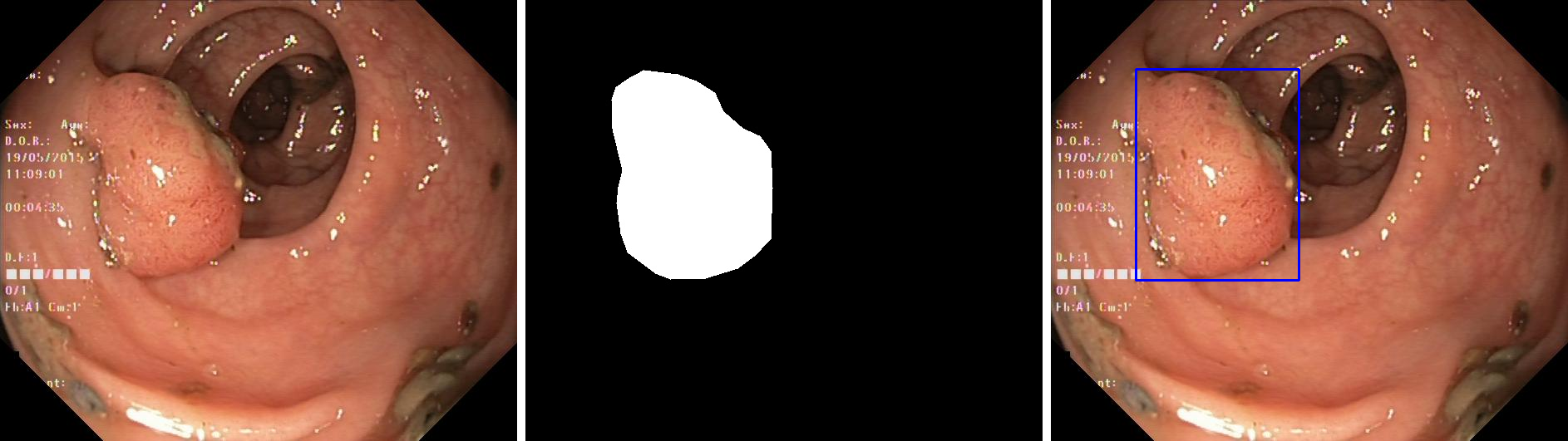}\\
     \vspace{0.5mm}
    \includegraphics[height=3cm]{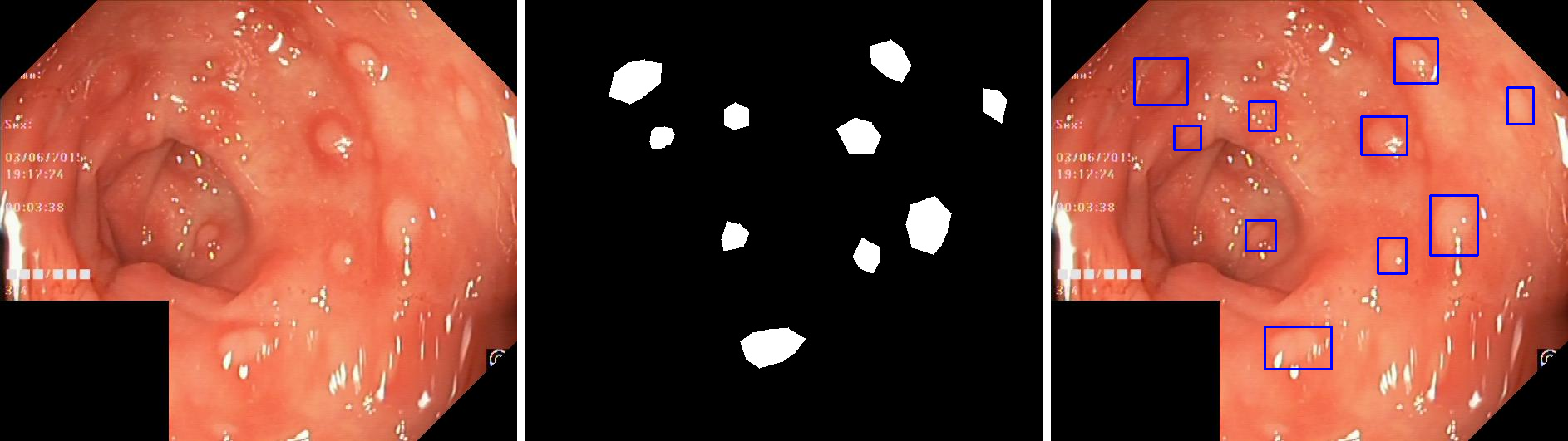}
    \caption{Examples of polyp images and their corresponding masks from Kvasir-SEG. The third image is generated from the original image using the bounding box information from the JSON file.}  
    \label{fig:images_and_bounding_box}
\end{figure}

The Kvasir dataset was used for the Multimedia for Medicine Challenge (the Medico Task) in 2017~\cite{riegler2017multimedia} and  2018~\cite{pogorelov2018medico} at the MediaEval Benchmarking Initiative for Multimedia Evaluation\footnote{http://www.multimediaeval.org} to develop and compare methods to reach clinical level performance on multiclass classification of endoscopic findings in the large bowel. However, the dataset was limited to frame classification only, due to only a frame-wise annotations. Thus, \citet{pozdeev2019automatic} trained their model on the CVC-ClinicDB, and tried to predict the segmentation masks for the Kvasir dataset, but could not report the experimental scores because of missing ground truth. 

\vspace{-2mm}
\subsection{The Kvasir-SEG Dataset Details}
To address the high incidence of colorectal cancer, we selected the polyp class of the Kvasir dataset for the initial investigation. The Kvasir-SEG dataset contains annotated polyp images and their corresponding masks. As shown in Figure~\ref{fig:images_and_bounding_box}, the pixels depicting polyp tissue, the \ac{ROI}, are represented by the foreground (white mask), while the background (in black) does not contain positive pixels. Some of the original images contain the image of the endoscope position marking probe from the ScopeGuide (Olympus).

The Kvasir-SEG dataset is made up of two folders: one for images and one for masks. Each folder contains 1000 images. The bounding boxes for the corresponding images are stored in a JSON file. Therefore, the kvasir-SEG dataset has image folder, masks folder and JSON file. The image and its corresponding mask have the same filename. The image files are encoded using JPEG compression, and online browsing is facilitated. The open-access dataset can be easily downloaded for research purposes at: https://datasets.simula.no/kvasir-seg/.  
\subsection{Mask Extraction}
We uploaded the entire Kvasir polyp class to Labelbox~\cite{Labelbox1} and created all the segmentations using this application. The Labelbox is a tool used for labelling the \ac{ROI} in image frames, i.e., the polyp regions for our case. A team consisting of an engineer and a medical doctor manually outlined the margins of all polyps in all 1000 images. The annotations were then reviewed by an experienced gastroenterologist.

Figure~\ref{fig:mask_extraction} shows example frames from the kvasir dataset where we have additionally marked the polyp tissue with green outline. After annotation, we exported the files to generate masks for each annotation. The exported JSON file contained all the information about the image and the coordinate points for generating the mask. To create a mask, we used \ac{ROI} coordinates to draw contours on an empty black image and fill the contours with white color. The generated masks are a 1-bit color depth images with white foreground and black background. Figure~\ref{fig:images_and_bounding_box} shows example images, their corresponding segmentation masks and bounding boxes from the Kvasir-SEG dataset. 
\section{Suggested Metrics}
\label{sec:Suggested_Metrics}
\vspace{-1mm}
Different metrics for evaluating and comparing the performance of the architectures exist.
For medical image segmentation tasks, the perhaps most commonly used metrics are Dice coefficient and \ac{IOU}. These are used in particular for several medical related Kaggle competitions~\cite{kagglemedicine}. In this medical image segmentation approach, each pixel of the image either belongs to a polyp or non-polyp region. We calculate the Dice coefficient and mean \ac{IOU} based on this principle. 

\textbf{Dice coefficient:} Dice coefficient is a standard metric for comparing the pixel-wise results between predicted segmentation and ground truth. 
It is defined as:
\begin{equation}
\text{Dice coefficient}(A,B) = \frac{2\times|A \cap B|}{|A| + |B|} =\frac{2 \times TP} {2 \times TP + FP + FN}
\end{equation}    
where A signifies the predicted set of pixels and B is the ground truth of the object to be found in the image. Here, TP represents true positive, FP represents false positive, and FN represents the false negative.

\textbf{Intersection over Union:}
The Intersection over Union (IoU) is another standard metric to evaluate a segmentation method. The IoU calculates the similarity between predicted (A) and its corresponding ground truth (B) as shown in the equation below:

\begin{equation}
IoU(A,B) =\frac{A \cap B} {A \cup B} = \frac{TP(t)} {TP(t) + FP(t) + FN(t)}
\end{equation}
In equation 2, t is the threshold. At each threshold value $t$, a precision value is calculated based on the above equation and parameters, which is done by calculating the predicted object to all the ground truth objects. There are other parameters such as recall, specificity, precision, and accuracy which are mostly used for frame-wise image classification tasks. The detailed explanation about these parameters can be found in the Kvasir dataset paper~\cite{kvasir}.
\section{Evaluation}
\label{sec:Baseline}
\label{sec:Application}
\vspace{-1mm}

The Kvasir-SEG dataset is intended for research and development of new and improved methods for segmentation, localization, and classification of polyps. To show that the dataset is useful for these purposes, we conducted several experiments, which we will described next.
\subsection{Baseline Models}

As our baseline, we have conducted initial investigations using two different methods. The first method is based on the efficient FCM~\cite{cai2007fast} unsupervised clustering algorithm. The second method is based on the deep-learning ResUNet~\cite{zhang2018road} architecture, utilizing the advantage of the residual block. 

When using basic \ac{CNN} architectures to predict outcomes in computer-vision tasks, millions of labelled training data are often needed to counteract overfitting and ensuring the model's ability to generalize when tested on new data~\cite{dravid2019employing}.
Because large datasets of medical images are hard to come by, using \acp{CNN} for medical-image segmentation systems remains
challenging. Image augmentation techniques~\cite{chollet2016building} and encoder-decoder architecture such as ResUNet~\cite{zhang2018road} are popular methods to use \acp{CNN} with smaller training sets.
\vspace{-2mm}
\subsection{Implementation Details}
Before applying the FCM algorithm, several pre-processing steps were applied to the dataset. First, we converted the image to grayscale and applied median blur to reduce noise. Then, we applied the Median-based Otsu method~\cite{otsu1979threshold}, which gave us the \ac{ROI}. Next, we converted image pixels between 0 and 1 and subtracted the image with its blurred version. We then used a threshold value and created an image with edges in it. Afterwards, we performed the dilation operation, which increases the foreground (white) region of the image. We subtracted edges from the image and clipped the image pixels value between 0 and 1. After that, we reshaped the image into 1D, which is the input to the FCM. Finally, the output of the FCM was reshaped into a 2D binary mask. 

For our experiment with the ResUNet model, we used image augmentation techniques like flipping, random crop, scaling, rotation, brightness, cutout, and random erasing to increase the size of our training dataset. After all pre-processing was completed, we resized our images to $320 \times 320$ pixels. We used 80\% of the dataset for training, and 10\% for validation. The remaining of 10\% was used for testing. We used five convolutional blocks both in the encoder and the decoder of the ResUNet model.
The batch size was set to 8, and we trained the model for 150 epochs. The proposed model converged at 91 epochs. We used a Nadam optimizer with the learning rate of 0.0001, $\beta$1 of 0.9 and $\beta$2 of 0.999. We chose Dice coefficient as the loss function and Relu as non-linearity. We used a threshold value t of 0.5 to convert the predicted masks pixels to foreground or background.  

For our deep-learning implementations, we used the Keras framework~\cite{chollet2018keras} and Tensorflow~\cite{abadi2016tensorflow} as a backend. We performed our experiment on a single Volta 100 GPU on a powerful Nvidia DGX-2~[1] AI system capable of 2\,PFLOPS tensor performance. The system is part of Simula Research Laboratories heterogeneous cluster and has dual Intel(R) Xeon(R) Platinum 8168 CPU@2.70\,GHz, 1.5\,TB of DDR4-2667\,MHz DRAM, 32\,TB of NVMe scratch space, and 16 of NVIDIAs latest Volta 100 GPGPUs~[2] interconnected using Nvidia's NVlink fully non-blocking crossbars switch capable of 2.4\,TB/s of bisectional bandwidth. The system was running Ubuntu 18.04.3\,LTS OS and had the latest Cuda 10.1.243 installed.
\vspace{-3mm}
\begin{table}[!t]
\vspace{-0.5cm}
\caption{Quantitative performance of ResUNet model on Kvasir-SEG dataset.}
\vspace{0.1cm}
\centering
\begin{tabular}{|p{2.5cm}||p{2.5cm}|p{2.5cm}|p{2.5cm}|}
\hline
\textbf{ResUNet} & \textbf{Loss} & \textbf{Dice coefficient} & \textbf{Mean IoU} \\ \hline
Train            & 0.059389      & 0.940609                & 0.920957          \\ \hline
Validation       & 0.196520      & 0.803479                 & 0.792339       \\ \hline
Test             & 0.212236      & 0.787763                  & 0.777771         \\ \hline
\end{tabular}

\label{table:table_1}
\end{table}
\begin{figure} [!t]
    \centering
    \includegraphics[width=10cm]{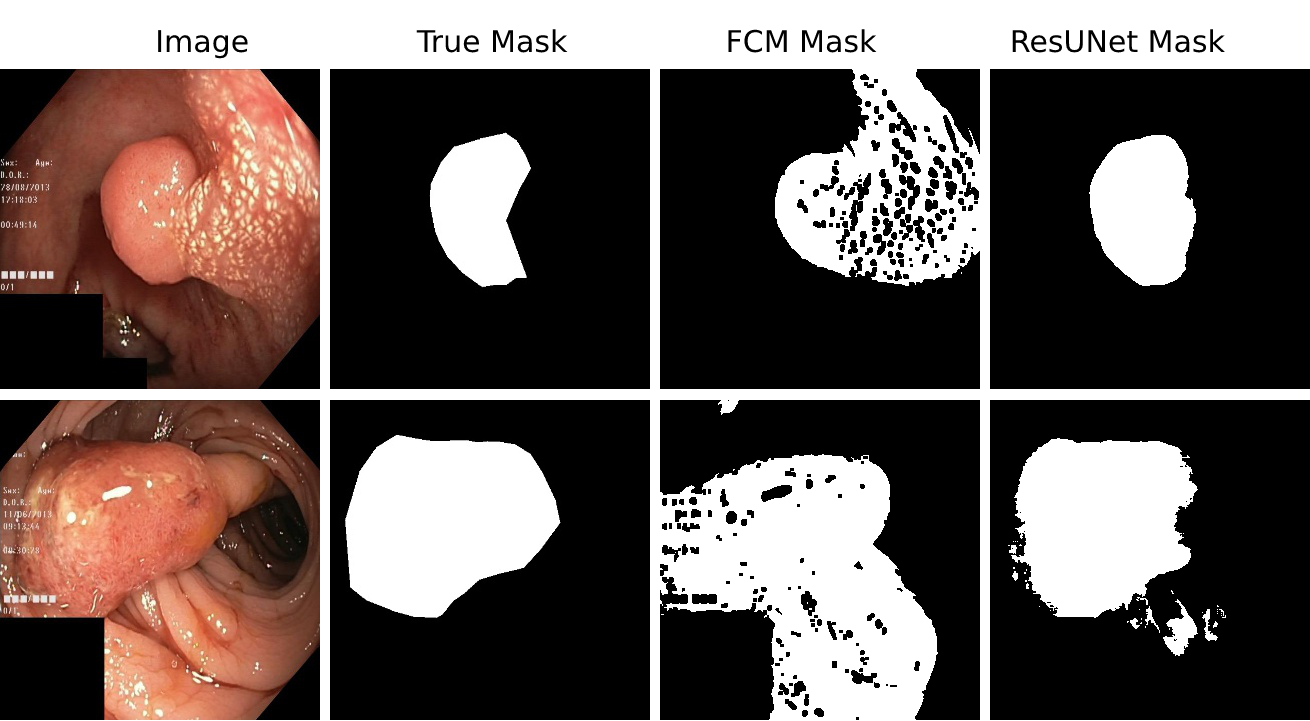}
    \caption{Qualitative comparison provided by both methods: Coloumn one shows the original image, column two shows the ground truth of the corresponding image. Column three shows the result of FCM clustering and column four shows the results of ResUNet.}
    \label{fig:results_comparison}
\end{figure}

\subsection{Results and Discussions}
The FCM clustering algorithm achieved a Dice coefficient of 0.239002 and a mean \ac{IOU} of 0.314187. The ResUNet model achieved a Dice coefficient of 0.787763 and mean \ac{IOU} of 0.777771 (see Table~\ref{table:table_1}) using the test dataset. We have included the training, validation and testing scores for the ResUNet model in Table~\ref{table:table_1}. Examples of qualitative result comparisons for the FCM algorithm and ResUNet model on the Kvasir-SEG dataset is shown in the Figure~\ref{fig:results_comparison}.  

\begin{figure} [!t]
    \centering
    \includegraphics[width=8cm]{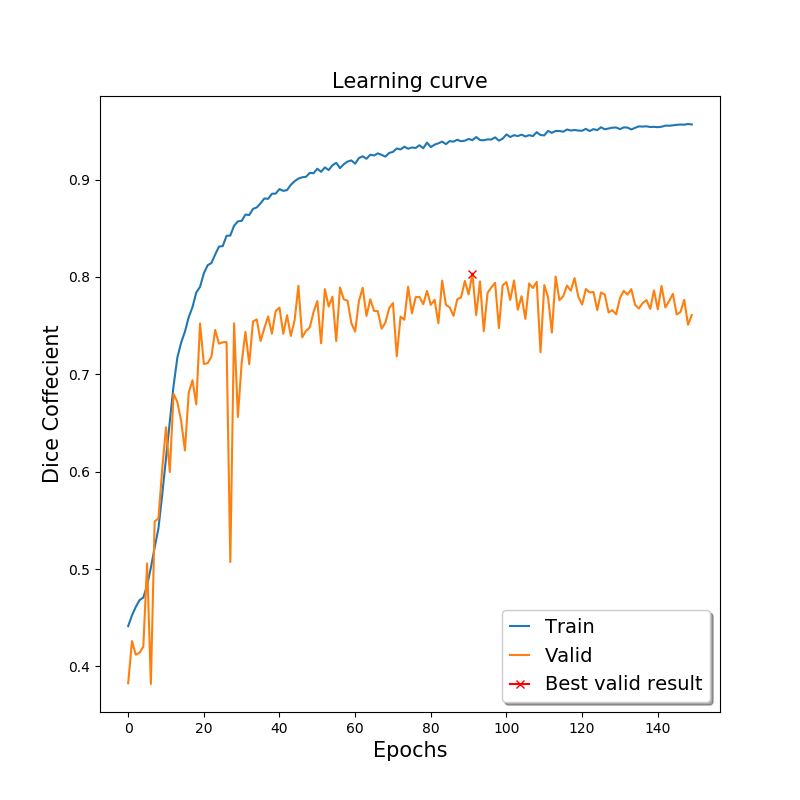}
    \caption{The learning curve of the proposed ResUNet model on Kvasir-SEG dataset showing Dice coefficient versus number of epochs. }
    \label{fig:learning_curve}
\end{figure}
Considering the quantitative and qualitative results (see Table~\ref{table:table_1} and Figure~\ref{fig:results_comparison}), the study shows a superior performance of the ResUNet model over the FCM algorithm in segmenting the polyp pixels. It should be noted that the FCM algorithm uses no data augmentation because it does not have any learning mechanism or learning parameters, whereas the ResUNet utilizes the advantage of the data augmentation techniques. Another important reason why FCM clustering approach did not perform well as it uses color as a significant feature for discriminating normal tissue and polyp. However, in practice, it is difficult to distinguish between polyps and other conditions inside the \ac{GI} tract on the basis of color features because of their similar appearances. We achieved promising results with the ResUNet model. 

There are no directly comparable papers with regards to our results. Nevertheless, compared to the work of Kang et al.~\cite{kang2019ensemble} which obtained the \ac{IOU} of 0.6607 and the work of~\citet{pozdeev2019automatic} that showed dice ranging from 0.6200 to 0.8600, we can say our results are either comparable or better. We think that the performance of our ResUNet model can be improved by providing it with more diverse polyp images. The plot of the learning curve of the ResUNet model is shown in Figure~\ref{fig:learning_curve}. The red mark in the learning curve \enquote{x} denote the best model. This best model was used for testing the previously unseen test dataset. The presented results are good; however, we believe that more research is required to achieve performance applicable to the clinic. 

\vspace{-2mm}
\section{Conclusion}
\label{sec:conclusion}
\vspace{-1mm}
In this paper, we present Kvasir-SEG: a new polyp segmentation dataset developed to aid multimedia researchers in carrying out extensive and reproducible research. We also present a FCM clustering algorithm and a ResUNet-based approach for automatic polyp segmentation. Our results show that the ResUNet model is outperforming the FCM clustering. 

The Kvasir-SEG dataset is released as open-source to the multimedia and medical research communities, hoping it can help evaluate and compare existing and future computer vision methods. This could boost the performance of computer vision methods, an important step towards building clinically acceptable \ac{CAI} methods for improved patient care. 
\vspace{-2mm}
\section*{Acknowledgements}
\vspace{-2mm}
This work is funded in part by the Research Council of Norway projects number 263248 (Privaton). We performed all computations in this paper on equipment provided by the Experimental Infrastructure for Exploration of Exascale Computing ($eX^3$), which is financially supported by the Research Council of Norway under contract 270053.

\bibliographystyle{splncsnat}
\bibliography{references.bib}
\end{document}